\documentclass[aps,pra,showpacs,twocolumn,groupedaddress]{revtex4}

\usepackage{graphics}

\begin{document}


\title{Global structure of Bose-Einstein condensates at high rotation:\\
Beyond the lowest Landau level description}


\author{Gentaro Watanabe$^{a,b,c}$ and C. J. Pethick$^{b}$}
\affiliation{
$^{a}$Department of Physics, University of Tokyo,
Tokyo 113-0033, Japan
\\
$^{b}$NORDITA, Blegdamsvej 17, DK-2100 Copenhagen \O, Denmark
\\
$^{c}$Computational Astrophysics Laboratory, RIKEN,
Saitama 351-0198, Japan}


\date{\today}

\begin{abstract}
We study the global density profile of
a rapidly rotating Bose-Einstein condensate in a harmonic trap
with transverse frequency $\omega_{\perp}$.
By introducing an additional variational
degree of freedom to the lowest Landau level wave function,
we demonstrate that with increasing strength of the interparticle interaction,
the global density profile changes from
a Gaussian to the inverted parabolic one characteristic of
Thomas-Fermi theory.
The criterion for the lowest Landau level wave function
to be a good approximation for the global structure is
that the mean field energy be small compared 
with  $\hbar \omega_{\perp}/N_{\rm v}$,
where $N_{\rm v}$ is the number of vortices in the cloud.
This condition is more stringent than the requirement that the mean field 
energy be small compared with  $\hbar \omega_{\perp}$
which is necessary for the lowest Landau level wave function
to be a good approximation to the local structure.
Our results show that the lowest Landau level wave function
is inappropriate for the global structure
of the system realized in recent experiments
even though this wave function can describe the local structure well.
\end{abstract}


\maketitle


Quantized vorticity and vortex lines are characteristic
features of superfluids \cite{donnelly,sonin}
and since the first observation of a vortex
in a Bose-Einstein condensed atomic gas \cite{first realization},
many beautiful experiments have been performed
to observe vortex lines and vortex lattices
in these systems \cite{creation,lattice1,lattice2}.
One fundamental question is how condensates behave when the vortex core
size become comparable to the spacing between vortices.  Such conditions
have not been achieved for superfluid liquid helium 4, but they have
been for atomic gases in harmonic traps.  Ho noted that the Hamiltonian
for a rotating gas in a harmonic trap is similar to that for charged
particles in a magnetic field, and he argued that for rotational angular
velocities just below the radial trap frequency $\omega_{\perp}$ all
particles would condense into the lowest Landau level (LLL) of the rotational
motion \cite{ho}.  Motivated by this suggestion, Schweikhard {\it et al.} have
recently achieved rotational angular velocities $\Omega$ in excess of
$0.99\omega_{\perp}$, at which the cloud contains a number of vortices
of order one hundred \cite{profile}.  The frontiers of experiments have
reached the so-called ``mean-field quantum Hall'' regime
where the $\hbar \Omega$ is large compared with the interaction energy, that is $\hbar \Omega\gg ng$,
where $n$ is the particle density and $g=4\pi
\hbar^2a/m$ is the effective two-body interaction, $m$ being the
particles mass and $a$ the scattering length.

Employing a quantum-Hall like wave function, which contains only
components in the lowest Landau levels, Ho predicted that the smoothed
density profile of a trapped cloud would be Gaussian.  To understand the
difference between the conventional approach to slowly-rotating
condensates and rapidly-rotating ones, the authors of Ref.\ \cite{bp}
(see also Ref.\ \cite{fb}) adopted a more general wave function consisting of
the product of a slowly varying envelope function, which describes the
global structure of the cloud and a rapidly varying one which describes
the properties of individual vortices.  They concluded that the density
profile would have the inverted parabolic form characteristic of
Thomas-Fermi theory, rather than the Gaussian one.

In the present paper, we consider the properties of a rotating cloud in
a potential $V(r)=m\omega_\perp^2 r^2/2$, using a trial wave function
which is a generalization of the lowest Landau level one.
We find that the global structure of the cloud is of the
Thomas-Fermi form provided $\hbar \Omega\gg ng/N_{\rm v}$,
where $N_{\rm v}$ is the number of vortices within the cloud.
For simplicity, we confine ourselves mainly
to the two-dimensional problem in this paper.

We shall use for the wave function $\Psi$ of the
condensed state the expression
\begin{equation}
\Psi({\bf r})=N^{1/2}h(r) \phi_{\rm LLL}
\equiv N^{1/2}\psi({\bf r})\ ,
\label{ansatz}
\end{equation}
where $N$ is the number of particles in the condensate and the
lowest Landau level wave function $\phi_{\rm LLL}$ has the general form
(i.e., quantum-Hall like wave function) as in Ref.\ \cite{ho}:
\begin{equation}
\phi_{\rm LLL}({\bf r})=A_{\phi}\ e^{-r^2/2
a_{\perp}^2}\ \prod_{i=1}^{N}(\zeta-\zeta_i)\ ,
\end{equation}
where $\zeta=x+iy$,
$\zeta_i$ are the vortex positions, $a_{\perp}=(\hbar/m\omega_{\perp})^{1/2}$ 
and $A_{\phi}$ is the normalization
constant.  The function $h$ which changes the density profile from the
form predicted by the lowest Landau level calculation,
is real, and varies slowly on the scale of the intervortex separation.
In the lowest Landau level wave function,  the positions of vortices
determine the density distribution,
apart from an overall multiplicative constant,
and the introduction of the modulating function
allows us to break this requirement.
The normalization condition for the wave function is
$\int d^2r h^2|\phi_{\rm LLL}|^2=1 $.
The normalization condition for $\phi_{\rm LLL}$ is
arbitrary, but for definiteness  we shall make the choice
$\int d^2r |\phi_{\rm LLL}|^2=1 $.

The energy per particle of the condensate in the rotating frame is given by
$E'=E-\Omega L_z$, where $E$ is the energy in the non-rotating frame and
$L_z$ is the expectation value of the angular momentum per particle about the axis of rotation.
Following Ho \cite{ho}, we write this in the form
\begin{eqnarray}
   &E'& = (\omega_{\perp}- \Omega) L_z \nonumber\\
&&\hspace{-1cm} + \int d^2r\ \psi^*
\left[\frac{m}{2}\left( \frac{\hbar\nabla_{\perp}}{im}
-\mbox{\boldmath $\omega$}_{\perp}\!\! \times \mbox{\boldmath $r$}\right)^2 
+ \frac{g_{\rm 2D}}{2} |\psi|^2\right] \psi\ ,
 \label{k original}
\end{eqnarray}
where $\mbox{\boldmath $\omega$}_{\perp}=  \omega_{\perp} \hat{\bf z}$.

An important observation is that
even if the spatial variations of $h$ are of order unity over the cloud,
the admixture of excited Landau levels in the wave function (\ref{ansatz})
is of order $a_{\perp}dh/dr \sim a_{\perp}/R \sim 1/N_{\rm v}^{1/2}$
relative to the LLL contribution \cite{footnote1}.
Here $R$ is the radial extent of the cloud and
$N_{\rm v} \sim R^2/a_{\perp}^2$ is the number of vortices in the cloud.
From this one can show that, apart from corrections of order
$\hbar(\omega_{\perp}-\Omega)/N_{\rm v}$,
for the wave function (\ref{ansatz}), one finds
\begin{eqnarray}
E'= \hbar \Omega +\int d^2r\ \langle|\phi_{\rm LLL}|^2 \rangle
\left\{
\frac{\hbar^2 }{2m}  \left(\frac{dh}{dr}\right)^2\right.\nonumber \\
\left.\vphantom{\left(\frac{dh}{dr}\right)^2}
+ \hbar (\omega_{\perp}-\Omega) \frac{r^2}{a_{\perp}^2}  h^2
+ \frac{bg_{\rm 2D}}{2}\langle |\phi_{\rm LLL}|^2 \rangle h^4
\right\}\ .\label{k}
\end{eqnarray}
Here $g_{\rm 2D}$ is the effective coupling parameter in two dimensions
and $b \equiv
\langle |\phi_{\rm LLL}|^4 \rangle / \langle |\phi_{\rm LLL}|^2 \rangle^2$
is a factor of order unity describing the renormalization
of the effective interaction due to the rapid density variations
on the scale of the vortex separation \cite{sinova, fb, bp}.

If the wave function is uniform
in the direction of the axis of rotation, $g_{\rm 2D}=Ng/Z$,
where $Z$ is the axial extent of the cloud, 
while if the wave function for motion in the direction of the rotation axis
corresponds to the ground state of a particle
in a harmonic oscillator potential
with frequency $\omega_{z}$, $g_{\rm 2D}=Ng/(\sqrt{2\pi}a_z)$,
where $a_z=(\hbar/m\omega_z)^{1/2}$ \cite{fetter2}.
Since the terms in braces in Eq.\ (\ref{k}) vary slowly in space,
we may obtain a good approximation to the energy
by making the averaged vortex approximation
as in Ref.\ \cite{ho}.
Like Ho, we shall in this paper assume that
the number of vortices per unit area is uniform.
If the vortex density is nonuniform,
this will increase the interaction energy,
which for a uniform system is a minimum for a regular triangular lattice.
For such a uniform array of vortices, one finds
\begin{equation}
\langle |\phi_{\rm LLL}|^2 \rangle
  \equiv \frac{1}{\pi \sigma^2} e^{-r^2/\sigma^2}.
\end{equation}
Here $\langle \ldots \rangle$ denotes an average over an
area of linear size large compared with the vortex separation
but small compared with $\sigma$.
The width $\sigma$ is given by $1/\sigma^2=1/a_{\perp}^2-\pi n_{\rm v}$,
where $n_{\rm v}$ is the average vortex density
in the plane perpendicular to the rotation axis \cite{ho}.

The first term in braces in Eq.\ (\ref{k}) represents the extra kinetic energy
associated with admixture of excited Landau levels.
It scales with the radius of the system
in the same way as the interaction energy.
If $g_{\rm 2D} \ll \hbar^2/m$,
the first term suppresses spatial variations of $h$,
and the solutions is $h\approx 1$,
and the wave function reduces to the LLL one.  
The condition for the LLL wave function to be a good approximation
may also be written as $Na/Z \ll 1$.
In current experiments $Na/Z$ is of order 10-100,
so this condition is strongly violated.
In the opposite limit, $g_{\rm 2D} \gg \hbar^2/m$,
the extra kinetic energy is unimportant,
and the optimal density profile is obtained by minimizing the second
and third terms in the integrand,
which results in a Thomas-Fermi density profile.
Another way to express the criterion for validity of the LLL approximation
is that the interaction energy per particle, $gn$, be small compared
with $\hbar \omega_{\perp}/N_{\rm v}$,
rather than the condition $\hbar \omega_{\perp}$
which has generally been assumed in earlier work.

To obtain quantitative results
we perform a variational calculation with a trial function that interpolates
between the Gaussian and Thomas-Fermi forms.
We exploit the fact that
$\lim_{\alpha\rightarrow\infty} (1-t/\alpha)^\alpha =e^{-t}$
as Fetter did in calculations for non-rotating clouds \cite{fetter}.
Thus we take
\begin{equation}
h(r)= A_h \left(1 - \frac{r^2}{\alpha L^2}\right)^{\frac{\alpha}{2}}
e^{r^2/2\sigma^2}\ ,\label{h gaussian-tf}
\end{equation}
for $0 \le r < \sqrt{\alpha}L$ and otherwise $h=0$.
The normalization factor $A_h$ is given by
$A_h^2=(\sigma^2/L^2)(1+1/\alpha)$.
The number of particles per unit area, divided by $N$,
is given by $\nu(r)=h^2 \langle |\phi_{\rm LLL}|^2 \rangle$
with the above expression (\ref{h gaussian-tf}) for $h$
can describe both Gaussian form ($\alpha \rightarrow \infty$)
and the Thomas-Fermi one ($\alpha =1$).
Hereafter, we shall refer to the expression (\ref{h gaussian-tf}) for $h$
as the Gaussian - Thomas-Fermi (G-TF) form.

Substituting Eq.\ (\ref{h gaussian-tf})
into Eq.\ (\ref{k}), we obtain the energy $E'_{\rm G{\mbox -}TF}$
for the Gaussian-TF modulation
\begin{eqnarray}
E'_{\rm G{\mbox -}TF} &=&
\hbar\Omega + \frac{\hbar\omega_{\perp}}{2} a_{\perp}^2
\left( \frac{1}{L^2}\frac{\alpha+1}{\alpha-1} - \frac{2}{\sigma^2}
+ \frac{\alpha}{\alpha+2}\frac{L^2}{\sigma^4} \right) \nonumber\\
&& + \hbar(\omega_{\perp}-\Omega)
\frac{\alpha}{\alpha+2}\frac{L^2}{a_{\perp}^2}
+ \frac{b g_{\rm 2D}}{2\pi}\frac{1}{L^2}\frac{(\alpha+1)^2}{\alpha(2\alpha+1)}\ .
\nonumber\label{energy g-tf}\\
\end{eqnarray}

Minimizing $E'_{\rm G{\mbox -}TF}$ with respect to $\sigma$ and $L$,
we obtain the following expressions for these two quantities:
\begin{eqnarray}
L^2 &=& a_{\perp}^2
\left(1-\frac{\Omega}{\omega_{\perp}}\right)^{-\frac{1}{2}} \nonumber\\
&& \times
\frac{1}{\alpha}\left[
(\alpha+2) \left\{ \frac{1}{\alpha-1}
+ \kappa \frac{(\alpha+1)^2}{(2\alpha+1)}
\right\} \right]^{\frac{1}{2}} ,\label{lsq}\\
\frac{1}{\sigma^2} &=&
\left( 1+\frac{2}{\alpha}\right) \frac{1}{L^2}
\ ,\label{sigmasq}
\end{eqnarray}
where
\begin{equation}
\kappa \equiv \frac{m b g_{\rm 2D}}{2\pi\hbar^2}\ ,
\label{kappa}
\end{equation}
is a dimensionless parameter determining
the strength of interparticle interactions.

Then we minimize $E'_{\rm G{\mbox -}TF}$ with respect to $\alpha$.
The optimal value of $\alpha$ is determined by the quartic equation
$\alpha^4 -2(1+4\lambda) \alpha^3 -12 \lambda\alpha^2 +2(1-3\lambda)\alpha
-1-\lambda=0$, where $\lambda\equiv\kappa^{-1}$.
Solving this equation for $\alpha$,
we find that the only physical solution is
\begin{widetext}
\begin{eqnarray}
\alpha &=& \frac{1}{2}(1 + 4\lambda)
+ \frac{1}{2}\sqrt{(1+4\lambda)^2+8\lambda-3\cdot 2^{\frac{2}{3}}
[\lambda(1+\lambda)]^{\frac{1}{3}}} \nonumber\\
&&+\frac{1}{2}
\left[
2\left\{(1+4\lambda)^2 +8\lambda\right\}
+ 3\cdot 2^{\frac{2}{3}} [\lambda(1+\lambda)]^{\frac{1}{3}}
+\frac{2(1+2\lambda)\left\{2\left[(1+4\lambda)^2+8\lambda\right]-3\right\}}
{\sqrt{(1+4\lambda)^2+8\lambda
-3\cdot 2^{\frac{2}{3}}[\lambda(1+\lambda)]^{\frac{1}{3}}}}
\right]^{\frac{1}{2}}\ ,\label{alpha}
\end{eqnarray}
\end{widetext}
since the three other solutions give negative or complex values of $\alpha$.
In the limits of weak $(\kappa\rightarrow 0)$
and strong interaction $(\kappa\rightarrow \infty)$ 
the above solution behaves as
\begin{eqnarray}
\alpha &\sim& 8/\kappa \qquad {\mbox (\kappa\ll 1)}\ ,\;\;\;{\rm and}
\label{alpha weak int}\\
\alpha &\simeq& 1+\frac{3}{2^\frac{1}{3}} \kappa^{-\frac{1}{3}}
+{\cal O}(\kappa^{-\frac{2}{3}})
\qquad {\mbox (\kappa^\frac{1}{3}\gg 1)}\label{alpha strong int}\ .
\end{eqnarray}
Observe that $\alpha$ does not depend on $\Omega$.

\begin{figure}[htbp]
\begin{center}\vspace{0.0cm}
\rotatebox{0}{\hspace{-0.cm}
\resizebox{7.5cm}{!}
{\includegraphics{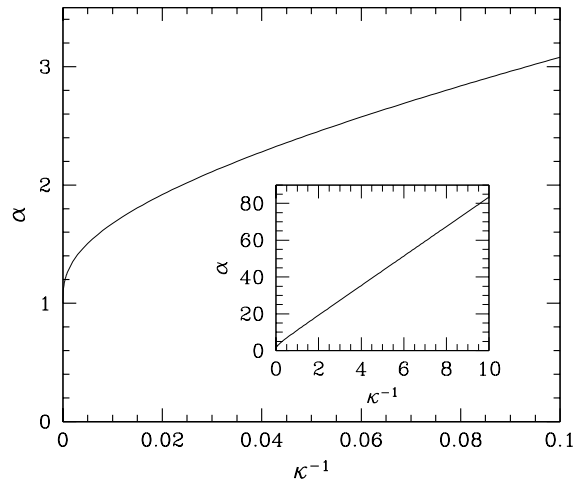}}}
\caption{\label{fig alpha}
  Shape index $\alpha$ as a function of
  $\kappa^{-1}$ covering a range of interaction strengths
  typical of current experiments ($\kappa\simeq100$).
  In the inset, $\alpha$ is plotted over a wider range of $\kappa^{-1}$
  to show its asymptotic behavior (\ref{alpha weak int}).
  }
\end{center}
\end{figure}

The shape index $\alpha$ given by Eq.\ (\ref{alpha}) is plotted
in Fig.\ \ref{fig alpha}, and it decreases from infinity
in the absence of interaction to unity as the
strength of the interaction ($\kappa$) increases from zero to infinity.
This means that the density profile of the cloud changes from
a Gaussian to an inverted parabola due to the interaction.
Without the minimization with respect to $\alpha$,
we can describe the cloud
with an assumption on the shape of the density profile
corresponding to the given value of the shape index.
For the parameters $L$ and $\sigma$, we can use the expressions
of Eqs.\ (\ref{lsq}) and (\ref{sigmasq}).
If we take $\alpha\rightarrow\infty$,
we reproduce a Gaussian density profile as in the case
without the modulating function,
but its width is optimized.

\begin{figure}[htbp]
\begin{center}\vspace{0.0cm}
\rotatebox{0}{\hspace{-0.cm}
\resizebox{8.0cm}{!}
{\includegraphics{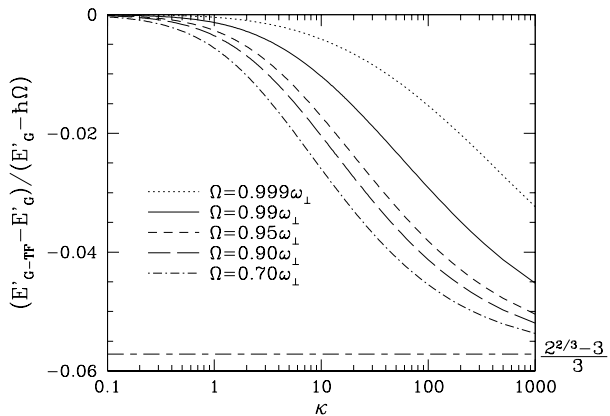}}}
\caption{\label{fig energy}
  Difference between the energies $E'_{\rm G{\mbox -}TF}$ calculated
  in the Gaussian-TF case with the optimized $\alpha$
  given by Eq.\ (\ref{alpha})
  and $E'_{\rm G}$ in the Gaussian case with $\alpha\rightarrow \infty$,
  relative to $E'_{\rm G}-\hbar\Omega$.
  For any value of $\Omega$, the curve converges to
  $(2^{3/2}-3)/3 \simeq -0.0572$
  in the limit of $\kappa\rightarrow \infty$.
  }
\end{center}
\end{figure}

Now we compare the results of the following two cases:
the Gaussian-TF profile with an optimized value of $\alpha$
and the Gaussian profile with $\alpha\rightarrow\infty$,
and discuss the situation relevant to the experiments.
For definiteness,
we write the energy for the Gaussian profile as $E'_{\rm G}$,
which can be obtained by taking $\alpha\rightarrow\infty$
in Eq.\ (\ref{energy g-tf}).
Relative energy differences
$(E'_{\rm G{\mbox -}TF}-E'_{\rm G})/(E'_{\rm G}- \hbar\Omega)$
between the Gaussian-TF and Gaussian cases are shown
in Fig.\ \ref{fig energy}.
In the limit $\kappa \rightarrow \infty$, the leading terms are
\begin{eqnarray}
E'_{\rm G} &\simeq& \hbar\Omega + \sqrt{2}\ \hbar\omega_{\perp}
\left(1-\frac{\Omega}{\omega_{\perp}}\right)^{\frac{1}{2}}
\kappa^{\frac{1}{2}}\ ,\\
E'_{\rm G{\mbox -} TF} &\simeq& \hbar\Omega
+ \frac{4}{3}\hbar\omega_{\perp}
\left(1-\frac{\Omega}{\omega_{\perp}}\right)^{\frac{1}{2}}
\kappa^{\frac{1}{2}}\ .
\end{eqnarray}
Thus, in this limit, the relative energy difference
$(E'_{\rm G{\mbox -}TF}-E'_{\rm G})/(E'_{\rm G} -\hbar\Omega)$
converges to $(2^{3/2}-3)/3\simeq -0.0572$
irrespective of the value of $\Omega$ (see Fig.\ \ref{fig energy}).
This result shows us that the density profile can change markedly
even though the energy reduction is less than 6\%.

We now consider the system realized in Ref.\ \cite{profile}
rotating at $\Omega\simeq 0.99\omega_{\perp}$, which is
typical of the highest angular velocities achieved so far.
If we assume that the density is uniform in the axial direction,
we get $\kappa=(2Na)/Z \agt 52$ using the relation $g_{\rm 2D}=g N/Z$
and the experimental data $Na/Z \agt 26$ given in Ref.\ \cite{profile}.
If we assume a Gaussian density profile in this direction
instead of the uniform one,
we obtain $\kappa=2Na/(\sqrt{2\pi} a_z)\simeq 1.4\times 10^2$
using $g_{\rm 2D}=g N/(\sqrt{2\pi}a_z)$, $\omega_z=2\pi\times5.3$ Hz,
$N=1.5\times10^5$ at $\Omega=0.989 \omega_{\perp}$, and
$a=5.6$ nm for the triplet state of $^{87}$Rb.
Thus we conclude that the typical value of the parameter $\kappa$
relevant to the experiments of Ref.\ \cite{profile} is of order 10$^2$.

\begin{figure}[htbp]
\begin{center}\vspace{0.0cm}
\rotatebox{0}{\hspace{-0.cm}
\resizebox{8.0cm}{!}
{\includegraphics{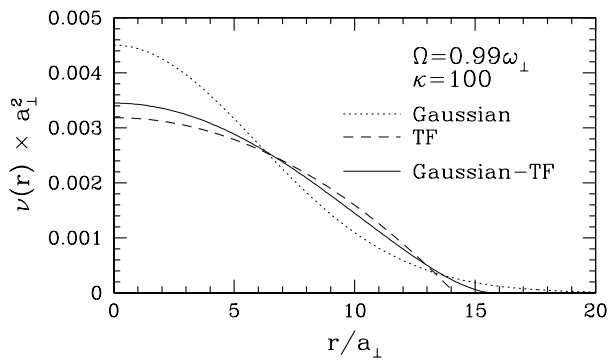}}}
\caption{\label{fig profile compare}
  Global density profile $\nu(r)=h^2 \langle |\phi_{\rm LLL}|^2 \rangle$
  of the cloud
  for $\Omega=0.99\omega_{\perp}$ and $\kappa=100$.
  The solid line is obtained for the optimized $\alpha$
  given by Eq.\ (\ref{alpha}). The dotted line and dashed one are
  obtained for $\alpha\rightarrow\infty$ and $\alpha=1$, respectively.
  }
\end{center}
\end{figure}

In Fig.\ \ref{fig profile compare},
we plot the global profile for the number of particles per unit area
$\nu(r)=h^2 \langle |\phi_{\rm LLL}|^2 \rangle$ at $\Omega=0.99\omega_{\perp}$
and $\kappa=100$ for the Gaussian-TF case in addition to
those for the two extreme cases of the Gaussian form
with $\alpha\rightarrow\infty$
and the inverted parabolic one labelled as ``TF''.
The latter is obtained by the procedure
corresponding to the Thomas-Fermi approximation in the non-rotating case
in which we neglect the contribution 
from the first term in the braces in Eq.\ (\ref{k})
[i.e., the second term in Eq.\ (\ref{energy g-tf})].
As can be seen from the solid line of this figure,
the density profile for the Gaussian-TF modulation is closer to the TF case
than to the Gaussian one and it can be fitted rather well by
an inverted parabola except at the edge of the cloud \cite{note fit}.
From Fig.\ \ref{fig energy} one can see that the relative energy reduction
compared with that for the Gaussian approximation is only $\simeq 2.9\%$.

In this paper, we have demonstrated that the global density profile
of a rapidly rotating Bose-Einstein condensate
is not well described by the lowest Landau level wave function,
even though the mean field energy is small
compared with $\hbar\omega_{\perp}$,
a condition satisfied in recent experiments.
By performing a variational calculation with a trial wave function
which includes components in excited Landau levels, we have shown 
how the density profile changes from a Gaussian to the Thomas-Fermi form
as the strength of the interparticle interaction increases.
The reduction in the energy resulting from the extra degree of freedom
in the wave function is of the order of one per cent. 
The approach described in this paper may be generalized
to time-dependent problems by deriving equations for the evolution
of the variational parameters from the condition that the action be stationary.

This work arose as a result of discussions and correspondence with Gordon Baym, to whom we are very grateful.
In addition, we thank V. Cheianov for helpful comments. G. W. thanks K. Sato, K. Yasuoka, and T. Ebisuzaki
for continuous encouragement and
L. M. Jensen, and P. Urkedal
for arranging the computer environment.
This work was supported
by Grants-in-Aid for Scientific Research
provided by the Ministry of
Education, Culture, Sports, Science and Technology
through Research Grant No. 14-7939.


\begin{thebibliography}{99}
\bibitem{note fit} In the region $r\alt 10a_{\perp}$,
the solid curve in Fig.\ \ref{fig profile compare}
can be fitted quite well by the inverted parabola 
$\nu(r) a_{\perp}^2=  0.00343 -2.1\times10^{-5} (r/a_{\perp})^2 $.
\end{thebibliography}

\end{document}